\documentclass[12pt,preprint]{aastex}

\shorttitle{VLBI Observations of the Seyfert 2 NGC 7674}
\shortauthors{Momjian et al.}

\begin{document}

\title{SENSITIVE VLBI CONTINUUM AND H~{\footnotesize I} ABSORPTION OBSERVATIONS
OF NGC 7674:\\
FIRST SCIENTIFIC OBSERVATIONS WITH THE COMBINED ARRAY VLBA, VLA and ARECIBO}

\author{Emmanuel Momjian\altaffilmark{1, 2}}
\email{momjian@pa.uky.edu}

\author{Jonathan D. Romney\altaffilmark{1}}
\email{jromney@nrao.edu}

\author{Christopher L. Carilli\altaffilmark{1}}
\email{ccarilli@nrao.edu}

\and

\author{Thomas H. Troland\altaffilmark{2}}
\email{troland@pa.uky.edu}

\altaffiltext{1}{National Radio Astronomy  Observatory, P O Box O, Socorro, NM
 87801.}
\altaffiltext{2}{University of Kentucky, Department of Physics and Astronomy,
 Lexington, KY 40506.}

\begin{abstract}

We present phase-referenced VLBI observations of the radio continuum emission
from, and the neutral hydrogen 21~cm absorption toward, the Luminous Infrared
Galaxy NGC~7674. The observations were carried out at 1380~MHz using the Very
Long Baseline Array, the phased Very Large Array, and the 305-m Arecibo radio
telescope. These observations constitute the first scientific use of the Arecibo
telescope in a VLBI observation with the VLBA. The high- and low-resolution
radio continuum images reveal several new continuum structures in the nuclear
region of this galaxy. At $\sim$100~mas resolution, we distinguish six continuum
structures extending over 1.4~arcsec (742~pc), with a total flux density of
138~mJy. Only three of these structures were known previously. All these
structures seem to be related to AGN activity.
The overall S-shaped pattern that the radio structures seem to form could be the result of the
interstellar medium diverting the outcoming jets from the central AGN. However,
we cannot rule out the possibility of a black hole merger that could result in a
similar structural pattern.
At the full resolution of the
array ($11 \times 5$~mas), we only detect two of the six continuum structures. Both are composed
of several compact components with brightness temperatures on the order of $10^{7}$~K. While it is
possible that one of these compact structures could host an AGN, they could also be shock-like
features formed by the interaction of the jet with compact interstellar clouds in the nuclear
region of this galaxy.
Complex H~{\footnotesize I} absorption is detected with our VLBI array at both high and low angular
resolution. Assuming that the widest H~{\footnotesize I} feature is associated with a rotating
H~{\footnotesize I} disk or torus feeding a central AGN, we estimate an enclosed dynamical mass
of $\sim$$7 \times 10^7~M{_\odot}$, comparable to the value derived from the
hidden broad H$\beta$ emission in this galaxy. The narrower H~{\footnotesize I} lines could
represent clumpy neutral hydrogen structures in the H~{\footnotesize I} torus.
The detection of H~{\footnotesize I} absorption toward some of the continuum components, and
its absence toward others, suggest an inclined H~{\footnotesize I} disk or torus in the central
region of NGC~7674. The overall averaged H~{\footnotesize I} spectrum toward the continuum
structures with H~{\footnotesize I} absorption is very consistent with the Arecibo single dish
H~{\footnotesize I} absorption spectrum at $3\rlap{.}'3$ resolution.

\end{abstract}


\section{INTRODUCTION}
At bolometric luminosities above $10^{11}~L{_\odot}$, infrared galaxies become the most
numerous objects in the local universe ($z \leq 0.3$). These
galaxies further subdivide into luminous (LIRGs\footnote{LIRG:
$10^{11} L{_\odot} \leq L{_{\rm IR}}(8-1000~{\mu}m) < 10^{12} L{_\odot}$}), ultraluminous
(ULIRGs\footnote{ULIRG: $10^{12} L{_\odot} \leq L{_{\rm IR}}(8-1000~{\mu}m) < 10^{13} L{_\odot}$}),
and hyperluminous (HyLIRGs\footnote{HyLIRG: $L{_{\rm IR}}(8-1000~{\mu}m) \geq 10^{13} L{_\odot}$})
infrared galaxies \citep{SM96}. A large fraction of these IR galaxies are interacting or merger
systems, and this fraction seems to increase with luminosity \citep{SAN88}. The bulk of the energy
radiated by these sources is infrared emission from warm dust grains heated by a central power
source or sources. The critical question concerning  these galaxies is whether the dust is heated by
a nuclear starburst, or an active galactic nucleus (AGN), or a combination of both.
Mid-infrared spectroscopic studies on a sample of ULIRGs by  \citet{GEN98},
suggest that 70\%--80\% of these galaxies are predominantly powered by recently
formed massive stars, and 20\%--30\% by a central AGN. These authors conclude
further that at least half of these ULIRGs are probably powered by both an AGN
and a starburst in a 1--2~kpc diameter circumnuclear disk or ring.

In this paper, we present VLBI continuum and H~{\footnotesize I} absorption
observations on the type-2 Seyfert galaxy NGC~7674 (Mrk~533, Arp~182,
UGC~12608). This galaxy is a nearby LIRG with an infrared luminosity of $L_{\rm
IR}= 3.1\times 10^{11} L{_\odot}$ \citep{DHL01}. Optical observations of
NGC~7674 show a nearly face-on spiral with an inclination angle of
$i=31^{\circ}$ \citep{VER97}. This galaxy is classified as Sbc pec \citep{VVC76}. It
is the brightest member of the well isolated Hickson~96 compact galaxy
group \citep{HIC82}, which consists of four interacting galaxies with a
mean heliocentric velocity of 8760~km~s$^{-1}$ and a low velocity dispersion
($\sigma_{\rm V}=160$~km~s$^{-1}$) \citep{VER97}. The two largest members in
this group, NGC~7674 (H96a) and NGC~7675 (H96b), are separated by a projected
distance of 2.2~arcmin. The observed features in NGC~7674, as revealed in
optical images, can be accounted for by tidal interactions with
its companion galaxies in the group \citep{VER97}.

The Very Large Array 15~GHz observations of NGC~7674 by \citet{UNG88} showed a
triple radio source with components defined as eastern, central, and western, but their 1666~MHz
EVN+MERLIN observations at 20~mas resolution only revealed the central and the
western components. These authors identified the central component as the radio
nucleus and the western component as a radio lobe moving outward through the
interstellar medium. However, their 20~mas VLBI image did not reveal any clear
jet structure connecting the central and the western components, which are
separated by $\sim$0.5~arcsec. MERLIN observations by \citet{BPM02} at
$0\rlap{.}''27 \times 0\rlap{.}''15$ and 1.4~GHz also revealed a triple radio
source consistent with the 15~GHz VLA observations \citep{UNG88}.

Single dish Arecibo observations by \citet{MIR82} at 3.3 arcmin resolution
showed two clearly separated H~{\footnotesize I} absorption features that are
blueshifted with respect to a wider emission line. The absorption lines have
full velocity widths at 20\% of 63 and 45~km~s$^{-1}$, while the emission line
has 344~km~s$^{-1}$. The two H~{\footnotesize I} absorption features are at
$\sim$8600~km~s$^{-1}$ and $\sim$8546~km~s$^{-1}$, respectively. However, the
neutral hydrogen interferometric observations with MERLIN at $0\rlap{.}''27
\times 0\rlap{.}''15$ and 32~km~s$^{-1}$ \citep{BPM02} revealed a single
H~{\footnotesize I} absorption feature against the central component with an optical depth of 0.21
and a velocity width of 114~km~s$^{-1}$ at half maximum, centered at
8623~km~s$^{-1}$. No H~{\footnotesize I} absorption was reported
toward the eastern or the western components in these MERLIN observations.

We adopt a distance of 115.6~Mpc to NGC~7674, assuming
${H_\circ=75}$~km~s$^{-1}$~Mpc$^{-1}$. At this distance 1~mas corresponds to
0.53~pc.

\section{OBSERVATIONS AND DATA REDUCTION}

The observations were carried out at 1380~MHz on September~15, 2001 using
NRAO's\footnote{The National Radio Astronomy Observatory is a facility of the
National Science Foundation operated under cooperative agreement by Associated
Universities, Inc.} Very long Baseline Array (VLBA) and the phased Very Large
Array (VLA), and NAIC's 305-m Arecibo radio telescope\footnote{The Arecibo
Observatory is part of the National Astronomy and Ionosphere Center, which is
operated by Cornell University under a cooperative agreement with the National
Science Foundation.}. These observations constitute the first scientific use of
the Arecibo radio telescope with the VLBA after the installation and
the successful testing of the VLBA4 recording system. Two adjacent 8~MHz baseband channel pairs were
used in these observations, both with right and left-hand circular polarizations, sampled at
two bits. The first baseband channel pair was centered close to the frequency of the
neutral hydrogen 21~cm line, which is at a heliocentric redshift of $z=0.0287$,
or $cz=8600$~km~s$^{-1}$. The velocity coverage of each baseband channel was
$\sim$1738~km~s$^{-1}$. The data were correlated at the VLBA correlator in
Socorro, NM with 512-point spectral resolution per baseband channel, and
5.1~seconds correlator integration time. The total observing time was 12~hours,
with the 305-m Arecibo radio telescope participating for 140 minutes. Table~1 summarizes the
parameters of these observations.

The inclusion of both the phased VLA (Y27) and Arecibo were essential in these
observations in two respects. One is the overall image
sensitivity (the rms thermal noise level), and the other is the short to moderate-length sensitive
baselines necessary to image large scale structures. While including the phased VLA in the
VLBA improves the image sensitivity by a factor of $\sim$2.4, the sensitivity
improves by a factor of $\sim$7.1 when both the phased VLA and Arecibo are added
to the VLBA. This latter sensitivity improvement is achieved only during the
time when Arecibo is observing. In our observations, where Arecibo participated
for about $20\%$ of the total observing time, the image sensitivity was
$\sim$4.2 times better than the VLBA alone for the whole observing time. Also,
the 305-m Arecibo radio telescope provided the 3rd sensitive short baseline in
our observations, this baseline is Arecibo--Saint Croix with 238~km
length. The other two sensitive short baselines are the Y27--Pie
Town (52~km) and Y27--Los Alamos (226~km). Short sensitive
baselines are critical to image extended continuum structures as well as the
H~{\footnotesize I} absorption toward these structures.

Data reduction and analysis were performed using the Astronomical Image
Processing System (AIPS) and the Astronomical Information Processing System
(AIPS++).

Along with the target source NGC~7674, the compact source J2329+0834 was
observed as a phase reference with a cycle time of 150 seconds, 100 seconds on
the target source and 50 seconds on the phase calibrator. The source J2334+0736
was used for amplitude calibration, and 3C~454.3 (J2253+1608) was observed as a
fringe finder and used to determine the initial fringe delays and to calibrate
the bandpass.

After applying {\it a priori} flagging and manually excising integrations
affected by interference, ionospheric corrections were applied using the AIPS
task ``TECOR''. Amplitude calibration was performed using measurements of
the antenna gain and the system temperature of each station. The delay solutions
for 3C~454.3 were applied and a bandpass calibration was performed. To image the
continuum of the target source, its line-free channels were
averaged to a single spectral channel with a total width of 14~MHz. The phase
calibrator J2329+0834 was self-calibrated in both amplitude and phase and imaged
in an iterative cycle. The self-calibration solutions of J2329+0834 were applied
on the continuum data of the target source NGC~7674. Due to the high rms noise
values obtained after applying the self-calibration solutions of the phase
reference source to NGC~7674, we performed self calibration on the continuum
data set of the target source itself. This resulted in a significant improvement
in the signal-to-noise ratio. The self calibration solutions of both the phase
calibrator and the target source were later applied to the line data, in which
every two channels were averaged to increase the sensitivity. The resulting
channel separation of the H~{\footnotesize I} data cube was 31.25~kHz
(6.986~km~s$^{-1}$). The continuum emission was subtracted from the
spectral-line visibility data using the AIPS task ``UVLSF''. The spectral-line
data were then analyzed at various spatial and spectral resolutions.
Deconvolution of the images in the continuum and line data sets was
performed using the Clark ``CLEAN'' algorithm as implemented in the AIPS task
``IMAGR''. Optical-depth $\tau(\nu)$ cubes were calculated from the
H~{\footnotesize I} absorption image cubes and continuum images with a similar
resolution as $\tau(\nu) = - {\rm ln} [ 1 - I_{\rm line}(\nu)/I_{\rm
continuum}]$.

Along with these observations, we also reduced and analyzed archival VLA data of
NGC~7674 from 1986. These A-array observations with program ID AH~233 were
carried out at 15~GHz.

\section{RESULTS AND ANALYSIS}

\subsection{{\it The Radio Continuum}}

Figure~1 shows moderate and high resolution continuum images of the nuclear region in NGC~7674
at 1380~MHz. The top image has a resolution of 20~mas (10.6~pc) obtained with
natural grid weighting (${\rm ROBUST}= 5$ in AIPS task ``IMAGR''). The resolution of this image is
comparable to the VLBI image of \citet{UNG88}. We distinguish six structures in this image, only
three of them previously known. We identify these three structures, following
\citet{UNG88}, as western (W), central (C) and eastern (E) components. In our
image, C is the brightest, E is the
diffuse emission region just to the east of C, and W is
the second brightest peak. In addition to these components, we detect a clear
collimated jet structure (J) connecting the C and W components,
and two diffuse emission regions, one located to the north-east of E
that we designate the northeastern component (NE), and another to the
south-west of W, the southwestern component (SW). The linear extent of the whole nuclear region in
the plane of the sky is $\sim$1.4~arcsec (742~pc), and the total flux is $\sim$138~mJy.
We also reduced the VLA data from these 1.38~GHz observations, which were performed in
the DnC configuration and yielded a resolution of $40\rlap{.}''5 \times
17\rlap{.}''3$ (PA 80${^\circ}$). The total flux density of the source at this
resolution is $\sim$~230~mJy. Our analysis of the 15~GHz archival VLA data resulted in a marginal
detection (3$\sigma$) of the newly discovered components NE and SW.

The bottom images in Figure~1 are of C ({\it left}) and W ({\it right}) at the
full resolution of our array, which is $11 \times 5$~mas ($5.8 \times 2.7$~pc). These two images
were obtained with grid weighting ${\rm ROBUST}= -1$. The other components identified in our 20~mas
image are resolved.

Figures~2 shows the continuum structure in gray scale with half-maximum ellipses for gaussians
fitted to the structures seen in C ({\it left}) and W ({\it right}).
These gaussian components provide a convenient measure of source structure even if they do not
necessarily represent discrete physical structures.
Parameters of the gaussian fits are listed in Tables~2 and 3, for components C and W, respectively.
Column~1 in both these tables lists the gaussians seen in Figure~2, and column~2 their positions
relative to the brightest peak in each image. Columns~3 and 4 show the peak and the total flux
densities of the fitted gaussian functions, respectively. Columns~5, 6, and 7 list
the half-power ellipse axes and the position angles. Column~8
shows the derived brightness temperatures of these components. All Gaussian fitting
parameters were obtained using the tool ``IMAGEFITTER'' in  AIPS++.

An image of the nuclear region of NGC~7674 at lower resolution (Figure~3) clearly
shows the diffuse components seen at 20~mas (Figure~1-top). The restoring beam
in Figure~3 is $92 \times 76$~mas ($48.8 \times 40.3$~pc). Table~4
summarizes the characteristics of the continuum structures labeled in  Figure~3.
The positions (col.~[2]) are relative to the central component (C).
Columns~3 and 4 show the peak and the total flux densities of the fitted gaussian functions,
respectively. Columns~5, 6, and 7 list the half-power ellipse axes and the position angles.
No good gaussian fit was obtainable for the jet structure; thus we have estimated its  flux density
and its size from the residual image left after subtracting the gaussian fittings of
the  other components. The reported size of the jet component is its linear
extent on the plane of the sky.

The accuracy of the phase calibrator position is important in phase-referencing
observations \citep {WAL99}, which allow the determination of the absolute
position of the target source and its components, if any, from the position of
the calibrator. The positions reported in all the above mentioned images and
tables are obtained from the position of the phase-reference source J2329+0834.
The position of J2329+0834 is from the VLBA calibrator survey \citep{BEA02},
with positional accuracy of $\sim$0.7~mas in right ascension and $\sim$1.3~mas
in declination.

Table~5 summarizes the magnetic field and pressure values of the radio
continuum components seen in NGC~7674 at low angular resolution using the
minimum energy condition \citep{MIL80}. These properties were derived using the
1380~MHz total flux densities of these components (Table~4) and their angular
extents on the plane of the sky, and assuming a path length through the source
on the line of sight equal to the major axis of the component. Spectral
indices of the C, E, and W components were obtained from \citet{BPM02}. For the
NE and SW components, we derived spectral indices from the marginal
detection of these structures at 15~GHz with the VLA and our 1380~MHz VLBI
results convolved to the beam size of the 15~GHz observations. Spectral indices of all five
components are steep ($>0.6;~S\propto\nu^{-\alpha}$). No spectral index estimate was possible for
the jet (J) structure, hence it was excluded from these calculations.

\subsection{{\it The} H~{\footnotesize I} {\it Absorption}}

All images reported in this section are reconstructed with natural
grid weighting in the AIPS task ``IMAGR''. The H~{\footnotesize I} data cube was
imaged using Hanning smoothing to improve the signal-to-noise ratio. Thus,
although the channel separation is about 7~km~s$^{-1}$, the effective velocity
resolution is $\sim$14~km~s$^{-1}$. The H~{\footnotesize I} results
are described separately for high and low spatial resolutions, because of the distinctive
characteristics of the absorption at these two scales.

\subsubsection{{\rm H~{\footnotesize I}} Absorption at High Spatial Resolution}

Figure~4 shows the naturally weighted high resolution continuum images of the
central ({\it top}) and the western ({\it bottom}) components in NGC~7674 with
averaged H~{\footnotesize I} absorption spectra against several background
continuum regions in both C and W. The angular resolution is $17 \times 5$~mas
($9 \times 2.7$~pc). The rms noise in the continuum images is $\sim$30~$\mu$Jy~beam$^{-1}$. While
the spectra obtained toward W show no H~{\footnotesize I} absorption, all the spectra obtained
against the components in C show significant absorption.
The rms noise in the H~{\footnotesize I} image cube is $\sigma \simeq $260~$\mu$Jy~beam$^{-1}$.

Figure~5 shows optical depth $\tau(\nu)$ images in the velocity range
8626.7--8529.0~km~s$^{-1}$. These images explicitly show the variation of the
H~{\footnotesize I} opacity against the central component of NGC~7674 at high
spatial resolution.
The images are obtained by blanking areas
where the flux density of the continuum image is below 2.5~mJy~beam$^{-1}$  (i.e.
less than 15\% of the peak flux), or the H~{\footnotesize I} absorption is
below 1.04~mJy~beam$^{-1}$ (i.e. 4~$\sigma$).

We distinguish four main absorption features in both the spectra H~{\footnotesize I} and the optical
depth images. They are identified in Figure~4. Their velocity widths range between 18 and
98~km~s$^{-1}$ at half maximum, with optical depths between 0.164 and 0.412. Table~6 summarizes
the physical characteristics of these four H~{\footnotesize I} absorption
features, obtained by fitting gaussian functions to the optical depth spectra
averaged over regions where the absorption features are seen. The fits were
performed using the tool ``IMAGEPROFILEFITTER'' in AIPS++. The velocities
(col.~[2]) refer to the center velocities of these features. The widths of these lines
(col.~[3]) are the full widths at half peak optical depth. Column~4 is the peak
optical depth value of each feature. $N_{\rm HI}/T_{\rm s}$ of each peak
(col.~[5]) is calculated as $N_{\rm HI}/T_{\rm s}~({\rm cm}^{-2}~{\rm K}^{-1}) =
1.823 \times 10^{18} \int \tau(v) dv$, with $\int \tau(v) dv= 1.06 \tau_{\rm
peak} \Delta v_{\rm FWHM}$ for gaussian profiles. The column densities $N_{\rm
HI}$ (col.~[6]) are derived assuming a spin temperature $T_{\rm s}$ of 100~K.

The first two H~{\footnotesize I} absorption components listed in Table~6 are
mainly seen against the continuum component C1. The weaker
of these two is wider, 98.3~km~s$^{-1}$. The other
two features are narrower, and are seen against the continuum component C3.
Only the strongest H~{\footnotesize I} absorption feature, the first component
in Table~7, has a measurable velocity gradient as shown in the velocity field
image with continuum contours superimposed (Figure~6-{\it top}). Figure~6-{\it
bottom} shows a position velocity (P-V) diagram along a cut in position angle
$14^{\circ}$. This diagram suggests a velocity gradient of
$\sim$30.3~km~s$^{-1}$ along $\sim$18.4~pc (i.e.
$\sim$1647~km~s$^{-1}$~kpc$^{-1}$).

Figures 7{\it a}-7{\it b} are $N_{\rm HI}/T_{\rm s}$ images in the velocity
range 8626.7--8529.0~km~s$^{-1}$. Figure~7{\it a} is a color-scale image and contours
of $N_{\rm HI}/T_{\rm s}$. Figure~7{\it b} shows the same $N_{\rm HI}/T_{\rm s}$
color-scale image with continuum contours superimposed for positional reference.

\subsubsection{{\rm H~{\footnotesize I}} Absorption at Low Spatial Resolution}

Figure~8 shows the naturally wighted low resolution continuum image of the
nuclear region in NGC~7674 at 1380~MHz and $129 \times 108$~mas ($68.4 \times
57.2$~pc) resolution. Averaged H~{\footnotesize I} absorption spectra are shown against the
previously identified continuum components. The rms noise level of the continuum and the
H~{\footnotesize I} data are 85 and 570~$\mu$Jy~beam$^{-1}$, respectively.
We notice clear H~{\footnotesize I} absorption detections (greater than
$3\sigma$=1.68~mJy~beam$^{-1}$) toward C, E, and NE. We do not
detect any absorption against SW, W, or J. The strongest H~{\footnotesize I}
feature is centered at $8600.3 \pm 1.1$. This value is consistent with the
strongest feature seen in the Arecibo single dish spectrum at $3\rlap{.}'3$
resolution \citep{MIR82}; however, it is lower by 23~km~s$^{-1}$ from the value
reported by \citet{BPM02} at $0\rlap{.}''27 \times 0\rlap{.}''15$ resolution
with MERLIN. The non-detection of H~{\footnotesize I} absorption in the
MERLIN observations toward E \citep{BPM02} could be simply a result of their
low velocity resolution, which is 32~km~s$^{-1}$, compared to 14~km~s$^{-1}$
achieved in our sensitive VLBI observations.

Figure~9 shows optical depth $\tau(\nu)$ images in the velocity range
8619.6--8522.0~km~s$^{-1}$. These images explicitly show the variation of the
H~{\footnotesize I} opacity against C, E, and NE at low spatial resolution. The
images are obtained by blanking areas
where the flux density of the continuum image is below 6~mJy~beam$^{-1}$  (i.e.
less than 12\% of the peak flux), or the H~{\footnotesize I} absorption is
below 2.28~mJy~beam$^{-1}$ (i.e. 4~$\sigma$).

Figures 10{\it a}-10{\it b} are $N_{\rm HI}/T_{\rm s}$ images in the velocity
range 8626.7--8522.0~km~s$^{-1}$. Figure~10{\it a} is a color-scale image and contours
of $N_{\rm HI}/T_{\rm s}$. Figure~10{\it b} shows the same $N_{\rm HI}/T_{\rm s}$
color-scale image with continuum contours superimposed for positional reference.

The top panel of Figure 11 is the velocity field of the H~{\footnotesize
I} absorption at low spatial resolution superimposed on the continuum contours,
and the bottom panel is the averaged H~{\footnotesize I} absorption
spectrum over the region where absorption signals are detected.

From the H~{\footnotesize I} absorption images and spectra at low resolution
(Figures 8, 9, 10, \& 11-{\it top}), we distinguish four main
absorption features. They are identified in Figure~8. Their velocity widths range between 23 and
165~km~s$^{-1}$ at half maximum, with optical depths between 0.1 and 0.65. Table~7 summarizes
the physical characteristics of these four H~{\footnotesize I} absorption
features, obtained by fitting gaussian functions to the optical depth spectra
averaged over regions where the absorption features are seen. The fittings were
performed using the tool ``IMAGEPROFILEFITTER'' in AIPS++. The velocities
(col.~[2]) refer to peaks of these features. The widths  of these lines
(col.~[3]) are the full widths at half peak optical depth. Column~4 is the peak
optical depth value of each feature as seen in the averaged spectra. $N_{\rm
HI}/T_{\rm s}$ of each peak (col.~[5]) is calculated as $N_{\rm HI}/T_{\rm
s}~({\rm cm}^{-2}~{\rm K}^{-1}) = 1.823 \times  10^{18} \int \tau(v)
dv$, with $\int \tau(v) dv= 1.06 \tau_{\rm peak} \Delta v_{\rm FWHM}$ for
gaussian profiles. The column densities $N_{\rm HI}$ (col.~[6]) are derived
assuming $T_{\rm s}=100~\rm  {K}$.

At this low resolution, the absorption feature with the highest velocity is seen
toward C. The velocity of the H~{\footnotesize I} line seen against E is
lower than the velocities of the H~{\footnotesize I} features seen against
C and NE.

The bottom panel of Figure~11 is the integrated H~{\footnotesize I} absorption
profile against the whole region where H~{\footnotesize I} signals are detected
at $129 \times 108$ mas resolution. While the high velocity feature is mainly
due to the H~{\footnotesize I} absorption toward C, the lower velocity feature
arises from the H~{\footnotesize I} absorption toward both E and NE. The
double peak pattern seen in our spectrum is very consistent with the observed
H~{\footnotesize I} absorption profile with Arecibo at $3\rlap{.}'3$ resolution
\citep{MIR82} in both velocity and width values. It should be noted that the
MERLIN observations of \citet{BPM02} did not detect the lower velocity component
seen in this spectrum.

\section{DISCUSSION}

\subsection{{\it The Radio Continuum}}

At moderate resolution (tens of mas), the nuclear region of NGC~7674 is
composed of several continuum structures extending over 0.75~kpc (Figure~1-{\it top}, \&  Figure~3).
At high resolution, the continuum results (Figure~1-{\it bottom}, \& Figure~2) show
several compact continuum structures in  both the central (C) and the western (W) components, with
brightness temperatures $\geq 10^{7}$~K. The total flux density of all the radio components seen in
our VLBI array is $\sim$138~mJy and represents only 60\% of the total flux
density seen at lower resolution with the VLA, and 72\% of the total flux density detected with
MERLIN \citep{BPM02}. These differences suggest the existence of diffuse
emission not detected by our VLBI array, which may be extensions to the continuum
structures seen in our observations. In the next two sections, we will discuss the possible nature
of all the continuum structures seen in this galaxy.

\subsubsection {Is there a starburst in NGC~7674?}
Most characteristics of the continuum structures in NGC~7674, both at large and small
scales, seem to be related to AGN activity and not to a nuclear or
circumnuclear startburst. Calculations by \citet{YRC01} clearly indicate that
this galaxy does not obey the radio-FIR correlation well known for ``normal''
galaxies, where the main energy source is not due to a supermassive  black hole
\citep{CON92}. This correlation is represented by the quantity $q$ as defined by
 \citet{HSR85}:
\begin{equation}
q={\rm log}~\{[ FIR/(3.75 \times 10^{12}]/S_{1.4~{\rm GHz}}\},
\end{equation}
where $FIR$ is given by:
\begin{equation}
FIR=1.26 \times 10^{-14}(2.58S_{60~{\mu{\rm m}}}+S_{100~{\mu{\rm m}}}).
\end{equation}
For ``normal galaxies'', $q=2.34\pm0.01$ \citep{YRC01}. Furthermore, these
authors conclude that a value of $q\leq1.64$ indicates the presence of an
AGN. For NGC~7674, they  report $q=1.53$.

Nonetheless, evidence does exist for star formation. The optical spectroscopic studies of
\citet{DHL01} indicate the  presence of young/intermediate-age (a few Myr to a few 100~Myr) stars in
the nuclear region of NGC~7674, and the question we seek to answer at this stage
is whether any of the emission structures we observe in the nuclear
region of NGC~7674 could be related to starburst activity. For this purpose, we
compare the expected star formation rates from H$\alpha$ and IR emissions
with the nonthermal radio continuum emission. The star formation rates will be
calculated using a Salpeter IMF which has the form $\psi(M)\propto M^{-2.35}$
\citep{SAL55} with mass limits 0.1 and 100~$M_\odot$ as given by \citet{KEN98},
taking into account that the nonthermal radio emission would result from Type
II supernova explosions that require stars with masses $\geq 8~M_\odot$.

The optical spectroscopic observations of \citet{GRI92} suggest that the
H$\alpha$ luminosity of the nuclear region of NGC~7674 is $5.3 \times
10^{34}$~W. Assuming all the H$\alpha$ emission is due to starburst activity
and using the relationship between the star formation rate (SFR) and the
H$\alpha$ luminosity \citep{KEN98}, which has the form:
\begin{equation}
SFR~(M_{\odot}~{\rm yr}^{-1})=7.9 \times 10^{-35}~L_{{\rm H}\alpha}~({\rm W}),
\end{equation}
we derive an SFR of 4.2~$M_{\odot}~{\rm yr}^{-1}$.

To calculate the SFR from the IR emission, we assume all the IR luminosity is
only due to starburst, and use the relationship given by \citet{KEN98}:
\begin{equation}
SFR~(M_{\odot}~{\rm yr}^{-1})=4.5 \times 10^{-44}~L_{\rm IR}~({\rm erg~s})^{-1}.
\end{equation}
The IR luminosity of NGC~7674 is $3.1\times 10^{11} L{_\odot}$ \citep{DHL01},
and the resulting SFR is 54~$M_{\odot}~{\rm yr}^{-1}$.

The SFR calculations based on radio luminosity will take into
account the flux densities of the three diffuse radio components, namely E, NE
and SW. The radio luminosity of these three components is $8.6 \times
10^{22}$~W~Hz$^{-1}$, and it is all due to nonthermal radiation because of the
mere fact that they are detected by our VLBI array. VLBI arrays, even with the extremes of
sensitivity we were able to achieve in these observations, currently lack the sensitivity to detect
thermal emission. To derive the SFR from the radio luminosity, we first calculate the supernova rate
using the empirical relation of \citet{CY90} that relates the observed Galactic
non-thermal luminosity ($L_{\rm NT}$) to the supernova rate $\nu_{\rm SN}$:
\begin{equation}
\Big(\frac{L_{\rm NT}}{10^{22}~{\rm W~Hz}^{-1}}\Big)=13
\Big(\frac{\nu}{\rm GHz}\Big)^{-0.8}
\Big(\frac{\nu_{\rm SN}}{{\rm yr}^{-1}}\Big).
\end{equation}
We derive $\nu_{\rm SN}$=0.86~yr$^{-1}$. The SFR can be calculated form the
supernova rate by using and combining the following expressions:
\begin{equation}
\nu_{\rm SN}=\int_{m_{\rm sn}}^{100M_\odot}~\psi(m)~dm,
\end{equation}
with $m_{\rm sn}=8~M_\odot$ for Type II supernovae, and
\begin{equation}
SFR=\int_{0.1M_\odot}^{100M_\odot}~m~\psi(m)~dm.
\end{equation}
These two equations result in an SFR of $\sim 116~M_{\odot}~{\rm yr}^{-1}$.

The very low SFR obtained from the H$\alpha$ emission could result from
extinction. However, the SFR estimated from the luminosity of the three diffuse
radio components in NGC~7674 is still greater than the SFR calculated from the
whole IR luminosity of the nuclear region.
Thus, we conclude that these continuum components are more likely related to
AGN activity, with no strong indication for a nuclear or
circumnuclear starburst in our VLBI results.

\subsubsection {The reason for the S-shaped pattern in NGC~7674, and the
location of the AGN}
Our continuum results at $\sim$100~mas resolution (Figure~3) revealed the previously
unknown continuum structures NE and SW. The overall shape of the continuum
components in the nuclear region seem to form an S-shaped pattern. A detailed
study on galaxies with similarly shaped radio structures was reported by \citet{ME02}.
The authors point out that the orientation of a black hole's spin axis could
dramatically change even in a minor merger, leading to a flip in the direction of any associated
jet. Thus, following their discussion, it
is possible that the components NE and SW are jet structures that formed
before a possible black hole merger in NGC~7674. However, it is also possible
that the interstellar medium could be diverting the outcoming jets from the AGN
and causing them to form the NE and SW structures and the overall S-shaped pattern we
are seeing in NGC~7674.

The black hole merger scenario would imply that the AGN may be located in the central component (C).
However, the few high angular resolution observations available on this source are not enough to
identify the AGN component in this galaxy. Early 18~cm VLBI observations at
20~mas resolution \citep{UNG88} tentatively identified the component
C to host the AGN, because of its brightness and proximity to the optical
center of the galaxy.
At a few mas resolution, the VLBA observations at 2.3 \& 8.4~GHz
(J.~F.~Gallimore \& C.~Murray-Krezan 2002, private communication) detected two
of the three compact  continuum features seen in the component C at 21~cm. These
components seem to match the C1 and C3 structures seen in our 21~cm
observations. The measured total fluxes at 2.3 \& 8.4~GHz were noisy, and the
derived spectral indices for the two components were neither flat, nor steep.
For $S\propto\nu^{-\alpha}$, these values were $\alpha(\rm C1)=0.53$ and
$\alpha(\rm  C3)=0.35$. Thus, it is not clear whether either of these two components is hosting the
AGN. Also, no information is available on the third structure seen in the central component. While
it is possible that any of these compact sources could be hosting the AGN, they could also be
shock-like features formed because of the interaction between the jet and
compact interstellar clouds in the nuclear region of NGC7674.

\subsection{{\it The} H~{\footnotesize I} {\it Absorption}}
Our observations reveal the details of the H~{\footnotesize I} absorption
against the radio continuum emission in NGC~7674 both at high angular resolution
($17 \times 5$~mas, Figures~4 to 7), and low resolution ($129 \times
108$~mas, Figures~8 to 11). The H~{\footnotesize I} absorption at high angular
resolution is composed of  several features. The strongest H~{\footnotesize I}
line is the only feature with a measurable  velocity gradient (Figure~6-top
panel). The P-V diagram in position angle 14$^{\circ}$  (Figure~6-bottom panel),
shows a velocity gradient of 1647~km~sec$^{-1}$~kpc$ ^{-1}$. If this measured
gradient is arising from H~{\footnotesize I} gas in a rotating disk or torus,
then at a radius of 17.4~mas (9.2~pc), assuming a Keplerian motion, the enclosed
dynamical mass would be {$5 \times 10^5~M{_\odot}$}. However, calculation based
on the observed width of the hidden broad H$\beta$ emission line
suggests a black hole mass in the range {$(4.4-13.2) \times 10^7~M{_\odot}$},
assuming an optical depth for electron scattering between $\tau_{\rm es}=0.1$
and 1 \citep{NT98}. This difference could be
explained by a rotating H~{\footnotesize I} disk, with only a small section of
it being illuminated by the background radio source. A similar explanation was
put forward by \citet{BPM02}. These authors further suggested that the rotating
H~{\footnotesize I} disk should be edge-on, because of the non-detection of
H~{\footnotesize I} absorption toward the western or the eastern components seen
on MERLIN scale.

However, our results clearly show H~{\footnotesize I} absorption against E and
NE, and no absorption against J, W, or SW (Figure~8). While a possible origin
for the H~{\footnotesize I} absorption toward E and NE could be neutral gas in
intervening clouds far from the nuclear region of NGC~7674, the almost
face-on orientation of the galaxy and the velocities of the detected H~{\footnotesize I}
features suggest that these H~{\footnotesize I} absorption lines also
arise within an H~{\footnotesize I} disk or torus. Thus, this rotating
H~{\footnotesize I} disk around the central black hole should be inclined,
i.~e.~should not be an edge-on structure.

Following the H~{\footnotesize I} torus model illustrated by \citet{P99} which suggests the
existence of dense neutral hydrogen clumps in the circumnuclear H~{\footnotesize I} torus
associated with a central AGN, it is very likely that the absorption features
seen at both low and high angular resolution with velocity widths
$<50$~km~s$^{-1}$ arise from such clumpy structures. Thus, the velocity
gradient seen in Figure~6 for the H~{\footnotesize I} feature with a width of
35.4~km~s$^{-1}$ at half maximum, could be due to turbulent motion or infall of
a dense clumpy structure within the H~{\footnotesize I} torus, and not
represent the rotational motion around the black hole. The absorption
seen toward E and NE components could also arise in such H~{\footnotesize
I} clumps within the torus. An indication for the clumpy
H~{\footnotesize I} torus can be seen from the optical depths and the velocity widths of
the various H~{\footnotesize I} components listed in Tables~6 and 7.
The narrower H~{\footnotesize I} features (i.e.~1, 3, 4, I, II, \& IV), which also have higher
optical depths ($\gtrsim 0.3$), seem to represent the neutral hydrogen clumps in the torus,
and the wider H~{\footnotesize I} components (i.e.~2 \& III), which have lower optical depths,
seem to represent the rotating H~{\footnotesize I} torus itself.
Using virial considerations, the enclosed dynamical mass estimated from the widest H~{\footnotesize
I} feature ($\Delta v_{\rm FWHM}=165$~km~s$^{-1}$) at a radius of about 10~pc is $\sim 7 \times
10^7~M{_\odot}$. This value is consistent with the dynamical mass derived from the hidden broad
H$\beta$ emission by \citet{NT98}.

\subsubsection{The Location of the AGN: a more comprehensive picture}
Observations of NGC~7674 with the BeppoSAX X-ray observatory by \citet {MAL98}
resulted in an absorption column density of $N_{\rm H} \simeq 6 \times
10^{22}~{\rm cm}^{-2}$, which is derived from the hard X-ray spectrum. Comparing
this column density with the values listed in column 5 of Tables~5 \& 6 would
suggest a spin temperature ($T_{\rm s}$) of at least 1000~K. Such a high value
could be expected for the H~{\footnotesize I} spin temperature if the absorption
is actually tracing a warmer atomic medium in the AGN environment \citep{GAL99},
where the spin temperature can rise to several 1000~K before the hydrogen atoms
are significantly ionized \citep{MHT96}.

However, the neutral hydrogen absorption studies in Seyfert galaxies by \citet{GAL99}
showed that the H~{\footnotesize I} absorption in these galaxies generally
avoids the nucleus, and it is more commonly seen against other radio components.
This suggests that the absorption in the X-ray could be tracing
different foreground gas than the absorption at 21~cm, and explains the
higher column density in X-ray compared to the 21~cm H~{\footnotesize I}
absorption. This also suggests that the continuum structures associated with the
H~{\footnotesize I} absorption, namely C, E, and NE, are jet structures. This possibility is further
supported by the steep spectra of these continuum structures seen at low resolutions discussed by
\citet{BPM02} and in section 3.1 of this work.

Based on the preceding discussion, we speculate that the
AGN is located to the west of all the continuum structures that have H~{\footnotesize I} absorption
associated with them. This picture would exclude the W and SW components as AGN hosts, because of
their steep spectra. Hence, a possible location for the AGN is in the J structure
revealed for the first time in these sensitive VLBI observations.
This implies that the two strongest sources, C and W, are possibly radio lobes
or hot spots, with the AGN located somewhere in between these two structures, but simply
free-free absorbed at 1.38~GHz.

\subsection{The age of the AGN in NGC~7674}

Using the above model for the location of the AGN and the nature of the C and W
components, we can estimate the age of the AGN in NGC~7674. For this purpose, we follow the
discussion in \citet{CB96} for non-relativistic high Mach jets. The standard assumption is that the
bulk of the kinetic energy supplied by the jet is L$_{\rm j}$=L$_{\rm R}/{\epsilon}$, where L$_{\rm
R}$ is the total radio lobe luminosity, and ${\epsilon}$ is an efficiency factor for converting bulk
kinetic energy into radio luminosity. The value of ${\epsilon}$ is $\leq0.4$, but it is more likely
to be considerably less (\citet{CB96} and references therein). Also, the jet must do
work (${\rm W} \approx {\rm P}_{\rm L} \times {\rm V}_{\rm L}$) to expand the
ambient medium, where P$_{\rm L}$ is the lobe pressure derived using the minimum
energy argument and V$_{\rm L}$ is the volume of the lobe. For the central
component in NGC~7674, L$_{\rm R}=8.5 \times 10^{40}~{\rm erg~s}^{-1}$, P$_{\rm
L} = 12.3 \times 10^{-9}~{\rm dyn~cm}^{-2}$, and V$_{\rm L}=6.84 \times
10^{61}~{\rm cm}^{3}$. The source lifetime t$_s$, which could be indicative of
the AGN lifetime, would be t$_s \approx {\rm W}/{\rm L}_{\rm j} \simeq
0.3/{\epsilon}$~Myr. Thus, the estimated age of the AGN in NGC~7674 is few Myr,
which is approximately the age of the young stellar population reported by
\citet{GEN98}.

\section{CONCLUSIONS}
We have presented the results of phase-referenced VLBI observations, using the
VLBA, the phased VLA, and the 305-m Arecibo radio telescope, of the 21~cm
continuum emission and the H~{\footnotesize I} absorption in the central
$\sim$0.75~kpc of the type-2 Seyfert galaxy NGC~7674, which is also classified as
a LIRG. The inclusion of both the phased VLA and Arecibo was a key factor in the
success of these observations.

The low resolution continuum images reveal previously unknown structures in the
nuclear region of this galaxy. The total VLBI flux density at 1380~MHz is 138~mJy,
and represents only 60\% of the total flux seen with the VLA at a lower
resolutions, suggesting the existence of diffuse emission not detected by our
VLBI array. All the observed structures seem to be related to AGN activity with
no strong indication of a nuclear or circumnuclear starburst. The overall 
S-shaped pattern that the radio structures seem to form could be the result of the
interstellar medium diverting the outcoming jets from the central AGN. However,
we cannot rule out the possibility of a black hole merger that could result in a
similar structural pattern.

The high resolution images show at least seven compact
sources in the components defined in the literature as central (C) and western
(W). Their brightness temperature is $\gtrsim 10^{7}$~K. These compact sources
are probably shock-like features formed due to the interaction of the jet
with interstellar gas. However, one of these compact
sources may actually be hosting the AGN. The currently available studies are
insufficient to confirm such a possibility. Future sensitive multi-frequency
VLBI observations will be the most direct way to identify a possible AGN
component in the nuclear region of NGC~7674.

Both low and high resolution VLBI images show a complex H~{\footnotesize I}
absorption with several peaks. The detected H~{\footnotesize I} lines have
column densities on the order of $10^{21}$~cm$^{-2}$, assuming a spin temperature of
100~K. At low resolution, we distinguish at least four H~{\footnotesize I}
absorption peaks detected against the central (C), the eastern (W), and the
northeastern (NE) components. No H~{\footnotesize I} absorption was seen against
the jet (J), the western (W), or the southwestern (SW) component. All these
absorption features seem to arise in the H~{\footnotesize I} disk or torus
associated with the AGN. While the narrow features ($\Delta v_{\rm FWHM}<50$~km~s$^{-1}$) could
represent clumpy neutral hydrogen structures in the H~{\footnotesize I} torus, the widest
absorption feature ($\Delta v_{\rm FWHM}=165$~km~s$^{-1}$) may represent
the rotating torus itself.
If the widest H~{\footnotesize I} feature (component III) is rotationally broadened by a massive
central object, the implied mass is about $10^{7}$~M${_\odot}$.
This value of the derived
enclosed mass is very consistent with the black hole mass calculated from
the width of the hidden broad H$\beta$ emission line. The detection of
H~{\footnotesize I} absorption toward some of the continuum components, and
its absence toward others, suggest that the H~{\footnotesize I} disk or torus is
inclined and not edge-on as previously had been suggested.

\section{ACKNOWLEDGMENTS}
The authors thank C. Salter and T. Ghosh for their contributions
to the success of the Arecibo part of these VLBI observations and their helpful
suggestions for processing the Arecibo data. This research has made use of the
NASA/IPAC Extragalactic Database (NED) which is operated by the Jet Propulsion
Laboratory, California Institute of Technology, under contract with the National
Aeronautics and Space Administration.
E.~M. is grateful for support from NRAO through the Pre-doctoral Research
Program. T.~H.~T. and E.~M. acknowledge NSF support through grant AST~99-88341.

\clearpage
\begin{figure}
\epsscale{0.9}
\plotone{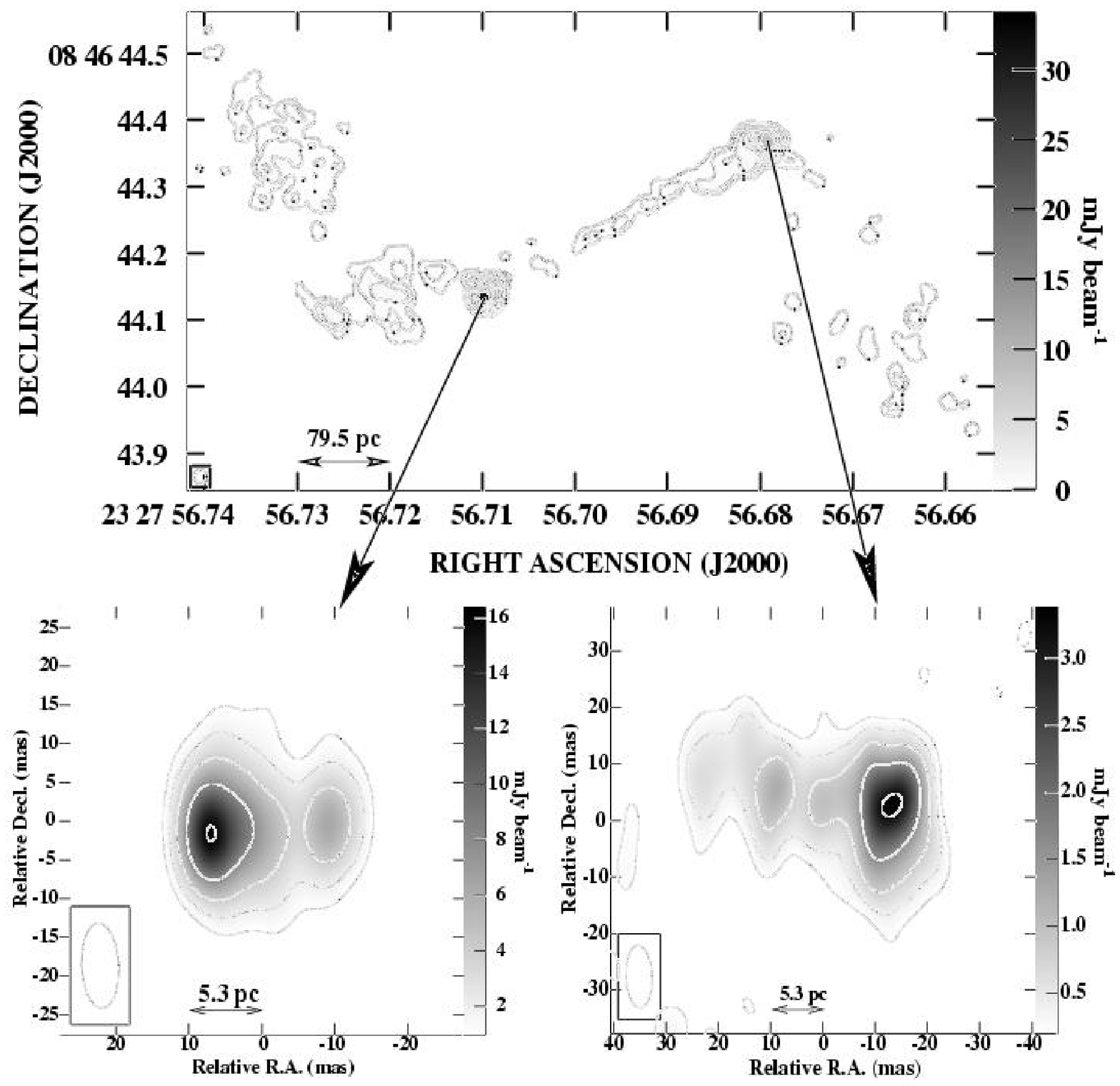}
\caption{{\it Top:} Continuum image of the nuclear region in NGC~7674 at
1380~MHz. A Gaussian taper falling to 30\% at 8~M$\lambda$ in the U direction
and 15~M$\lambda$ in the V direction was applied. The restoring beam size is
20~mas. The peak flux is 34~mJy~beam$^{-1}$, and the contour levels
are at $-$5, 5, 10, 20, 40,$\ldots$640 times the rms noise level, which is
34~$\mu$Jy~beam$^{-1}$. {\it Bottom:} Full resolution continuum images of the
central ({\it left}) and the western ({\it right}) components in NGC~7674. The
restoring beam size is $11 \times 5$~mas in position angle $3^{\circ}$. The
peak fluxes are indicated with the wedge on the right side of each image. The
rms noise in both images is 67~$\mu$Jy~beam$^{-1}$.
The contour levels of the bottom-left image are at $-$1, 1, 2, 4, 8,
16~mJy~beam$^{-1}$, and the (0,0) point is $\alpha \rm{(J2000)}=23^{\rm h}
27^{\rm m} 56\rlap{.}^{\rm s} 7097$, $\delta \rm{(J2000)}=08^{\circ} 46'
44\rlap{.}'' 135$. The contour levels of the bottom-right image are at
$-$0.2, 0.2, 0.4, 0.8, 1.6, 3.2~mJy~beam$^{-1}$, and the (0,0) point is $\alpha
\rm{(J2000)}=23^{\rm h} 27^{\rm m} 56\rlap{.}^{\rm s} 6798$, $\delta
\rm{(J2000)}=08^{\circ} 46' 44\rlap{.}'' 367$.
\label{FIG1}}
\end{figure}

\clearpage
\begin{figure}
\epsscale{0.9}
\plotone{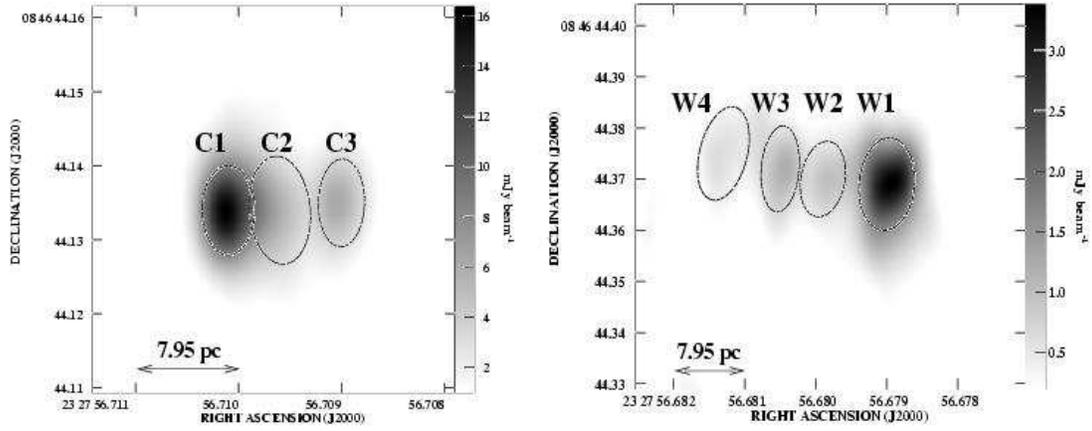}
\caption{Gray scale continuum image of the central ({\it left}) and western
 ({\it right})
components in NGC~7674 at $11~\times~5$~mas resolution. The ellipses represent
the half-power gaussians fitted to these components.
\label{FIG2}}
\end{figure}

\clearpage
\begin{figure}
\epsscale{0.9}
\plotone{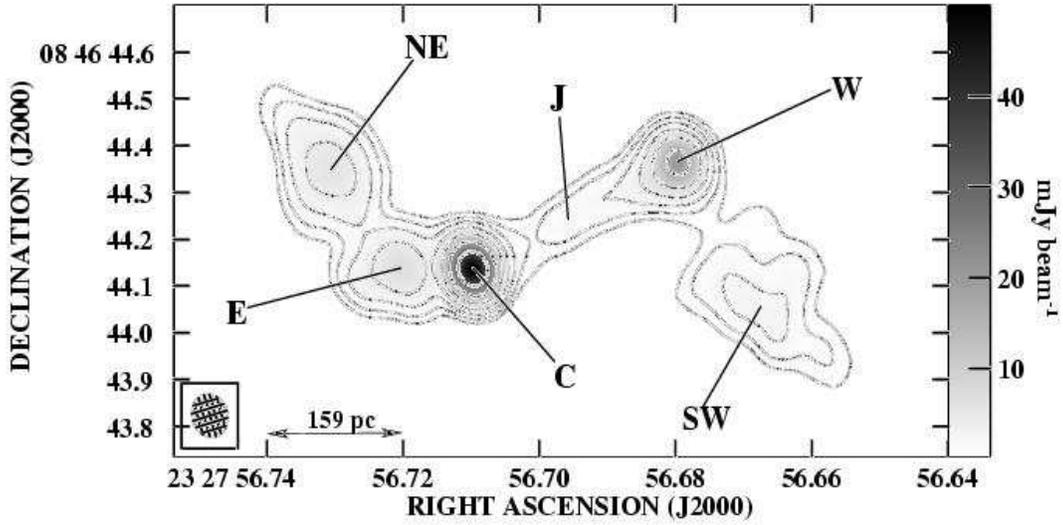}
\caption{Continuum image of the nuclear region in NGC~7674 at
1380~MHz. The restoring beam size is $92 \times 76$~mas in position angle
$16.5^{\circ}$. A two dimensional Gaussian taper falling to 30\% at
2.5~M$\lambda$ was applied. The contour levels are $-$0.5, 0.5, 1, 2, 4 ,8,
16, 32~mJy~beam$^{-1}$. The rms noise level is 78~$\mu$Jy~beam$^{-1}$.
The properties of the labeled components are listed in Table~4.
\label{FIG3}}
\end{figure}

\clearpage
\begin{figure}
\epsscale{0.9}
\plotone{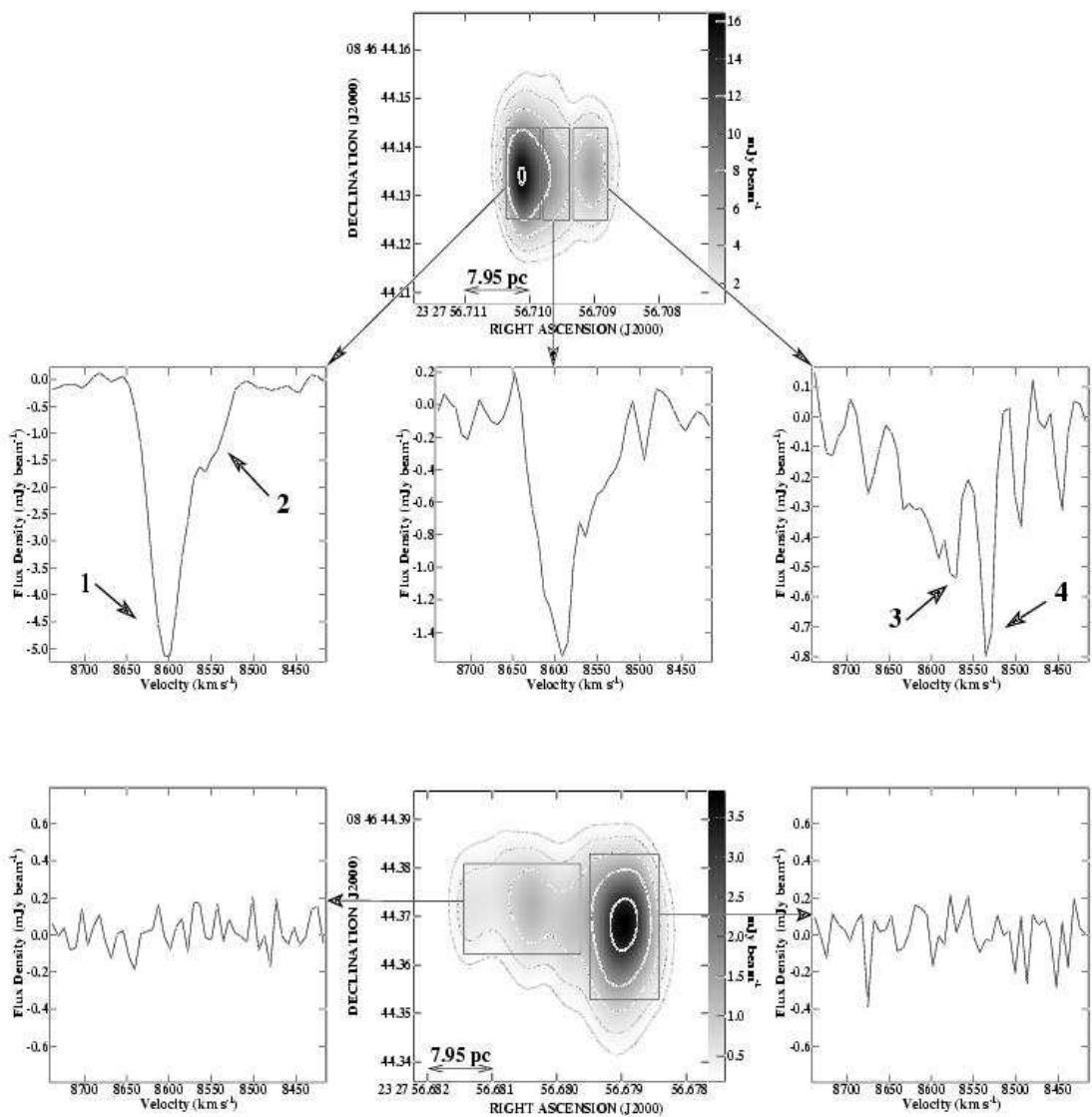}
\caption {H~{\footnotesize I} absorption spectra obtained
at various locations against the central ({\it top}) and the western ({\it
bottom}) components of NGC~7674 at 1380~MHz. The effective velocity resolution
of the H~{\footnotesize I} spectra is 14~km~s$^{-1}$. The restoring beam size
of both the continuum and the H~{\footnotesize I} is $17 \times 5$~mas in
position angle $3^{\circ}$. The contour levels of the continuum images are the
same as in Figure~1-{\it bottom}.
\label{FIG4}}
\end{figure}

\clearpage
\begin{figure}
\epsscale{0.9}
\plotone{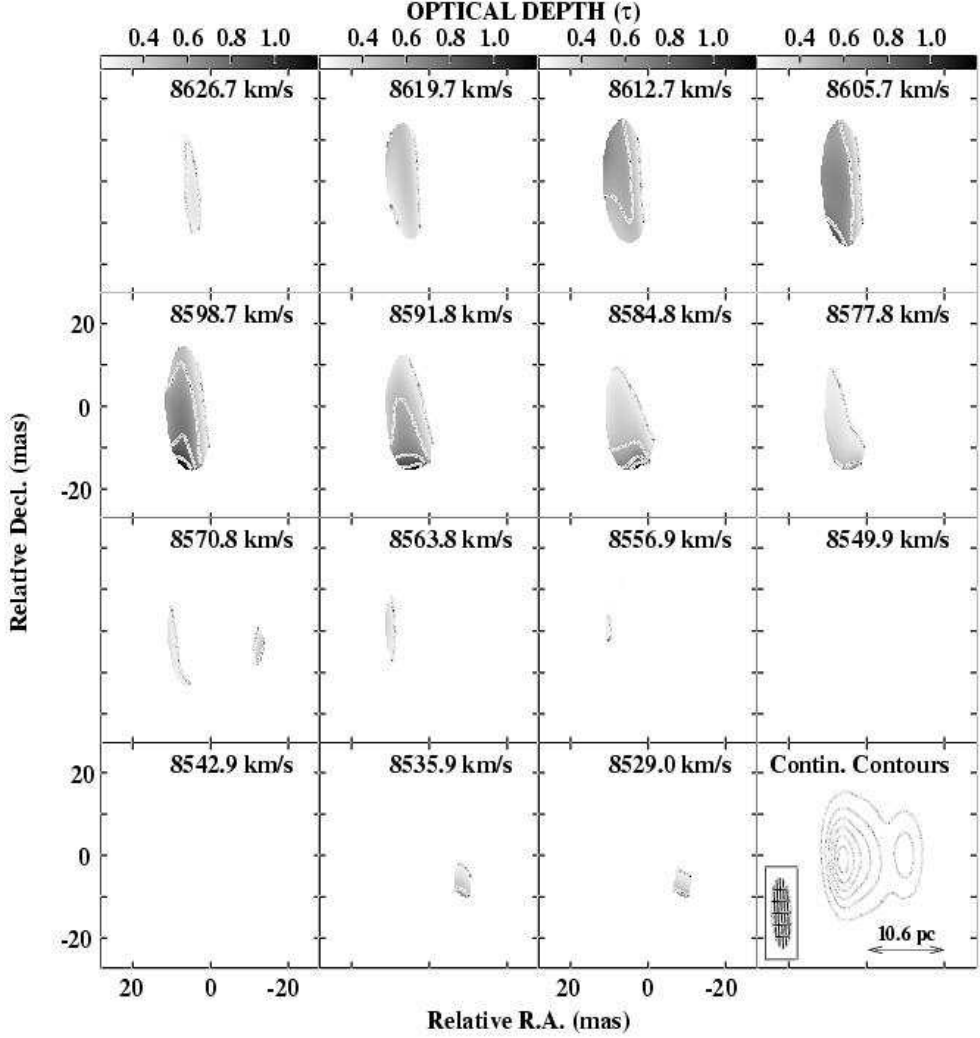}
\caption {High resolution gray-scale and contour H~{\footnotesize I} optical
depth channel images toward the central component in NGC~7674 in the velocity
range 8626.7--8529.0~km~s$^{-1}$. The gray-scale range is indicated by the step
wedge at the top of the images; the contour levels are 0.25, 0.5, 0.75,
and 1.0. At the bottom-right corner a contour image of the continuum
is shown as a positional reference with levels at 2.5, 5, 7.5, 10, 12.5 and
15~mJy~beam$^{-1}$. The restoring beam in these images is $17
\times 5$ in position angle $3^{\circ}$. The point (0,0) is $\alpha
\rm{(J2000)}=23^{\rm h} 27^{\rm m} 56\rlap{.}^{\rm s} 7097$, $\delta
\rm{(J2000)}=08^{\circ} 46' 44\rlap{.}'' 135$.
\label{FIG5}}
\end{figure}

\clearpage
\begin{figure}
\epsscale{0.8}
\plotone{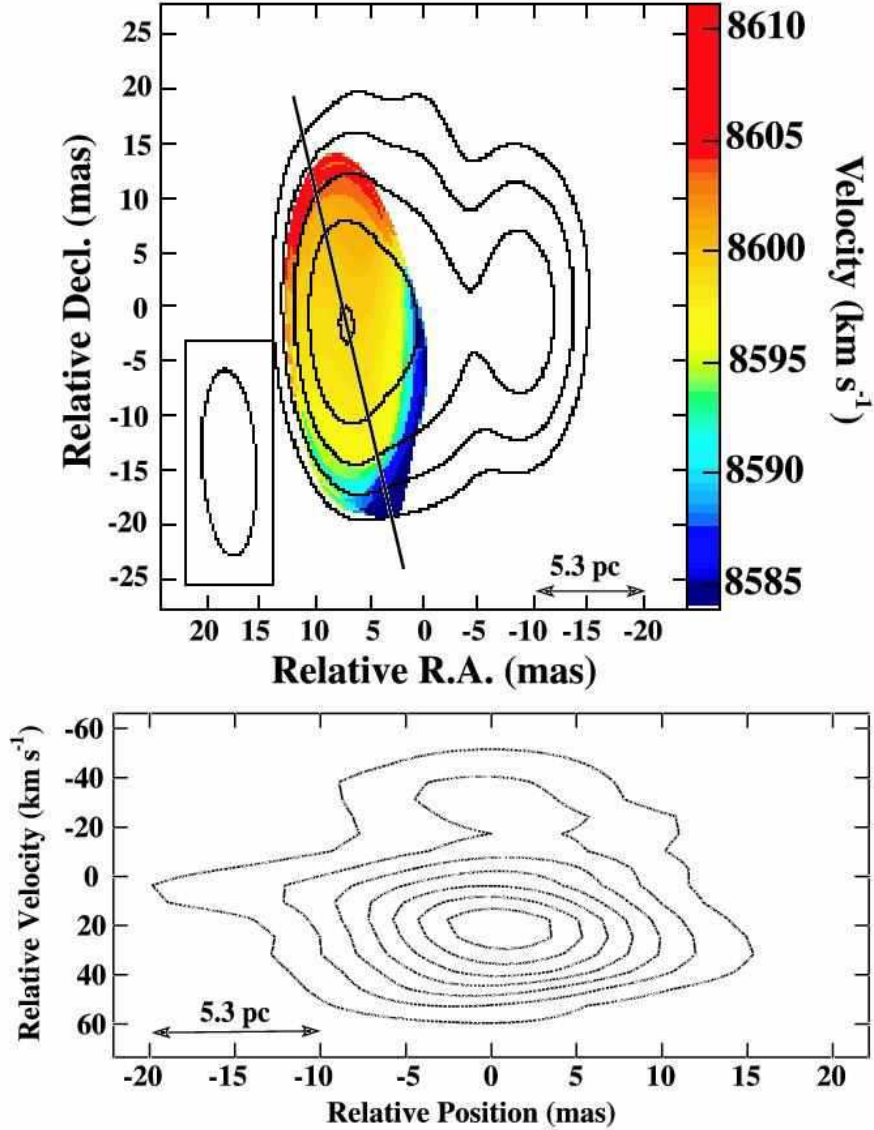}
\caption {{\it Top}: Velocity field of the strongest H~{\footnotesize I}
feature at high spatial resolution, superimposed on the continuum contours with
levels at 1, 2, 4, 8, 16~mJy~beam$^{-1}$. The $17 \times 5$ mas
(P.A.~3$^{\circ}$) restoring beam is shown in the lower left corner. The point
(0,0) is $\alpha \rm{(J2000)}=23^{\rm h} 27^{\rm m} 56\rlap{.}^{\rm s} 7097$,
$\delta \rm{(J2000)}=08^{\circ} 46'44\rlap{.}''135$.
{\it Bottom}: Position-velocity plot of the 21~cm H~{\footnotesize I} absorption along a cut in
position angle 14$^{\circ}$ as shown in the top panel image. The zero point on the velocity scale
corresponds to a heliocentric velocity ($cz$) of 8601.1~km~s$^{-1}$. The contour levels of the P-V
plot are at (-4, -8, -12,$\ldots$ -28) $\times$ 1$\sigma$ (260~$\mu$Jy~beam$^{-1}$).
\label{FIG6}}
\end{figure}

\clearpage
\begin{figure}
\epsscale{0.55}
\plotone{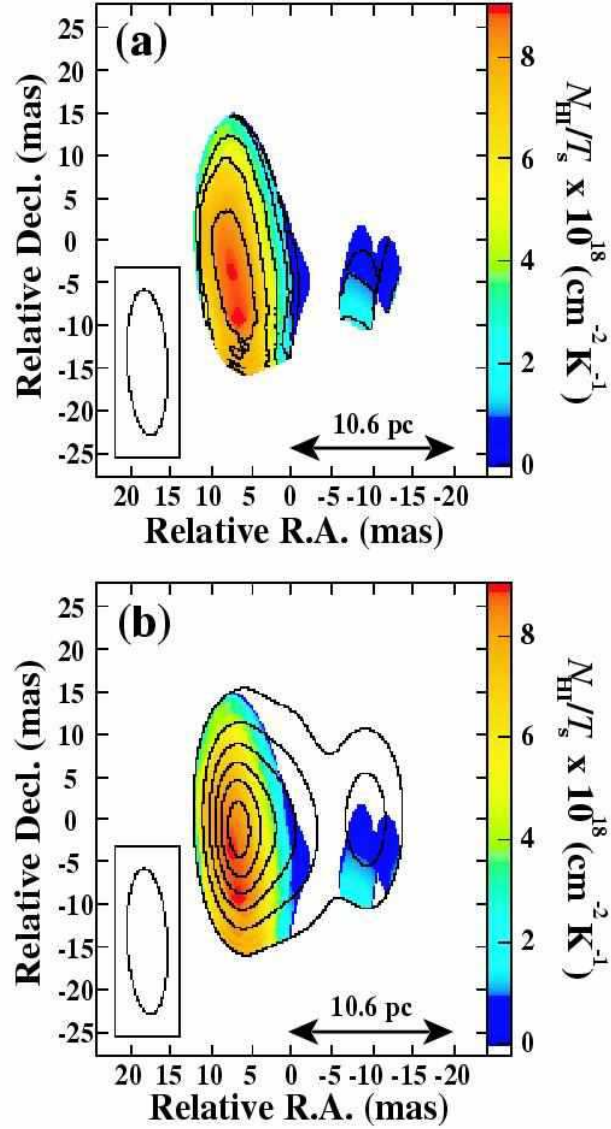}
\caption {$N_{\rm HI}/T_{\rm s}$ images in the velocity range
8626.7--8529.0~km~s$^{-1}$: (a) $N_{\rm HI}/T_{\rm s}$ color image and
contours superimposed; the contours are at 0.5, 1, 2, 4, 6, 8 $\times$
10$^{18}$~cm$^{-2}$~K$^{-1}$ (b) $N_{\rm HI}/T_{\rm s}$ color image with
continuum contours superimposed; the contour levels are 2.5, 5, 7.5, $\ldots$
15~mJy~beam$^{-1}$. The $17 \times 5$ mas (P.A.~3$^{\circ}$) restoring beam
is shown in the lower left corner of each image. The point (0,0) is
$\alpha \rm{(J2000)}=23^{\rm h} 27^{\rm m} 56\rlap{.}^{\rm s} 7097$, $\delta
\rm{(J2000)}=08^{\circ} 46'44\rlap{.}''135$.
\label{FIG7}}
\end{figure}

\clearpage
\begin{figure}
\epsscale{0.75}
\plotone{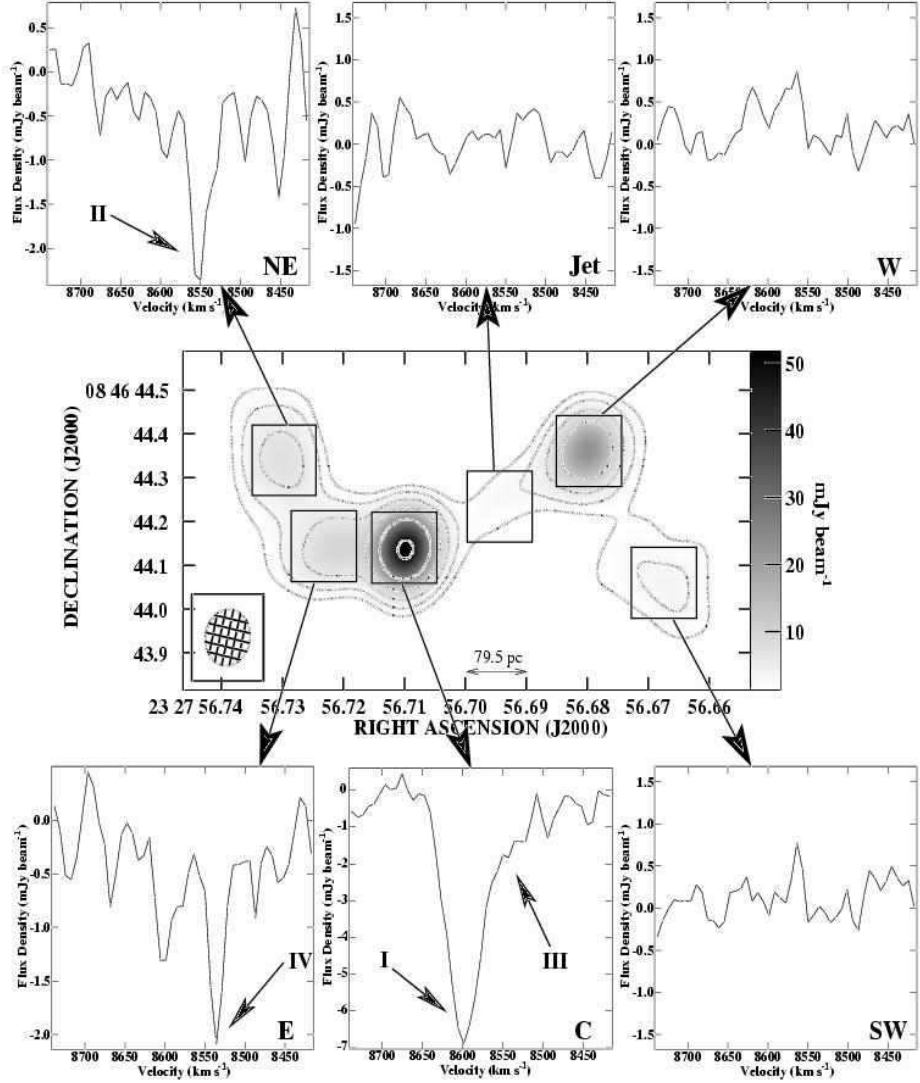}
\caption {H~{\footnotesize I} absorption spectra averaged against the
continuum components of NGC~7674 as seen at low spatial resolution at
1380~MHz. The effective velocity resolution of the H~{\footnotesize I} spectra
is 14~km~s$^{-1}$. The restoring beam size of both the continuum and the
H~{\footnotesize I} is $129 \times 108$~mas in position angle $-13^{\circ}$. The
contour levels of the continuum image are at $-$1.5, 1.5, 3,
6,$\ldots$48~mJy~beam$^{-1}$. The rms noise level of the continuum and the
H~{\footnotesize I} data are 85 and 570~$\mu$Jy~beam$^{-1}$, respectively.
A two dimensional Gaussian taper falling to 30\% at 1.4~M$\lambda$ was applied
to obtain these images.
\label{FIG8}}
\end{figure}

\clearpage
\begin{figure}
\epsscale{0.8}
\plotone{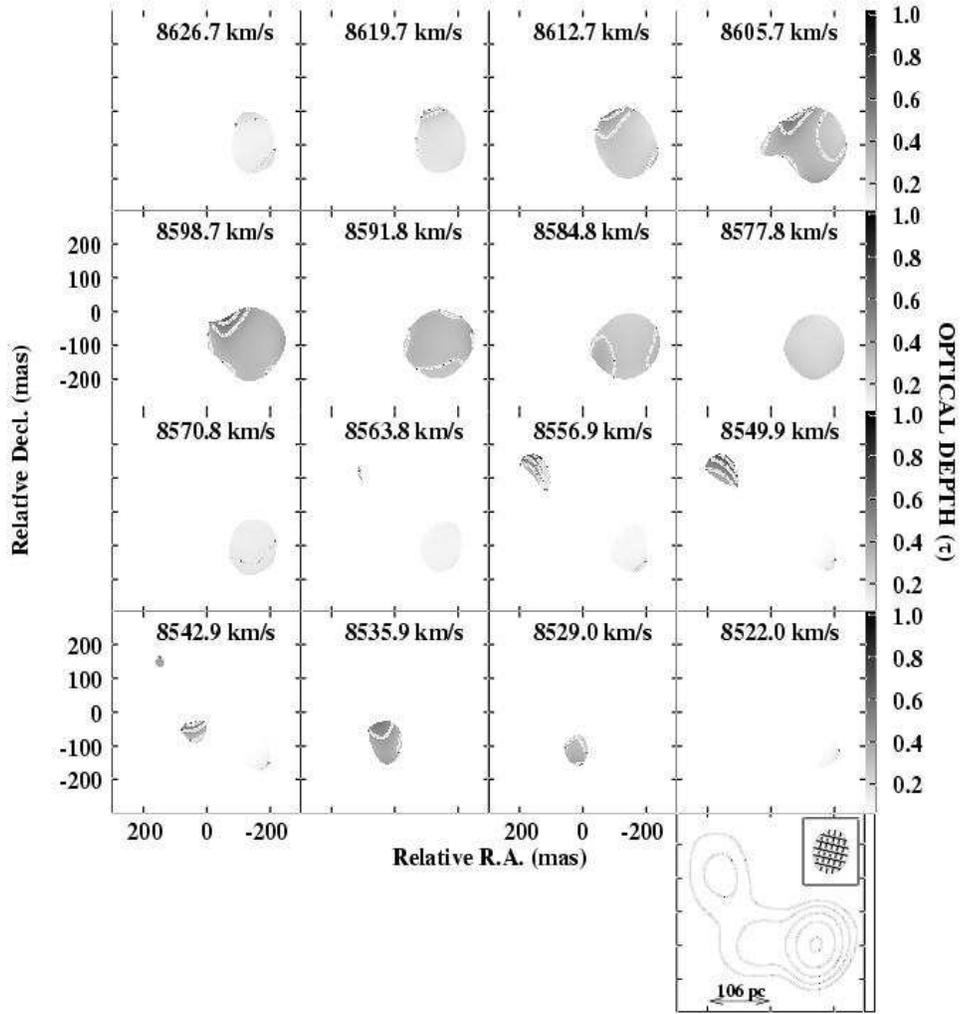}
\caption {Low resolution gray-scale and contour H~{\footnotesize I} optical
depth channel images toward the central, eastern, and northeastern components
in NGC~7674 in the velocity range 8626.7--8522.0~km~s$^{-1}$. The gray-scale
range is indicated by the step wedge on the right-side of the images; the
contour levels are 0.15, 0.3, 0.45, 0.6, 0.75 and 0.9. At the bottom-right corner a
continuum image is shown as a positional reference with levels at
3, 6, 12, 24 and 48~mJy~beam$^{-1}$. The restoring beam in these
images is $129 \times 108$ in position angle $13^{\circ}$. The point (0,0) is
$\alpha \rm{(J2000)}=23^{\rm h} 27^{\rm m} 56\rlap{.}^{\rm s} 7195$, $\delta
\rm{(J2000)}=08^{\circ} 46' 44\rlap{.}'' 234$. \label{FIG9}} \end{figure}

\clearpage
\begin{figure}
\epsscale{0.6}
\plotone{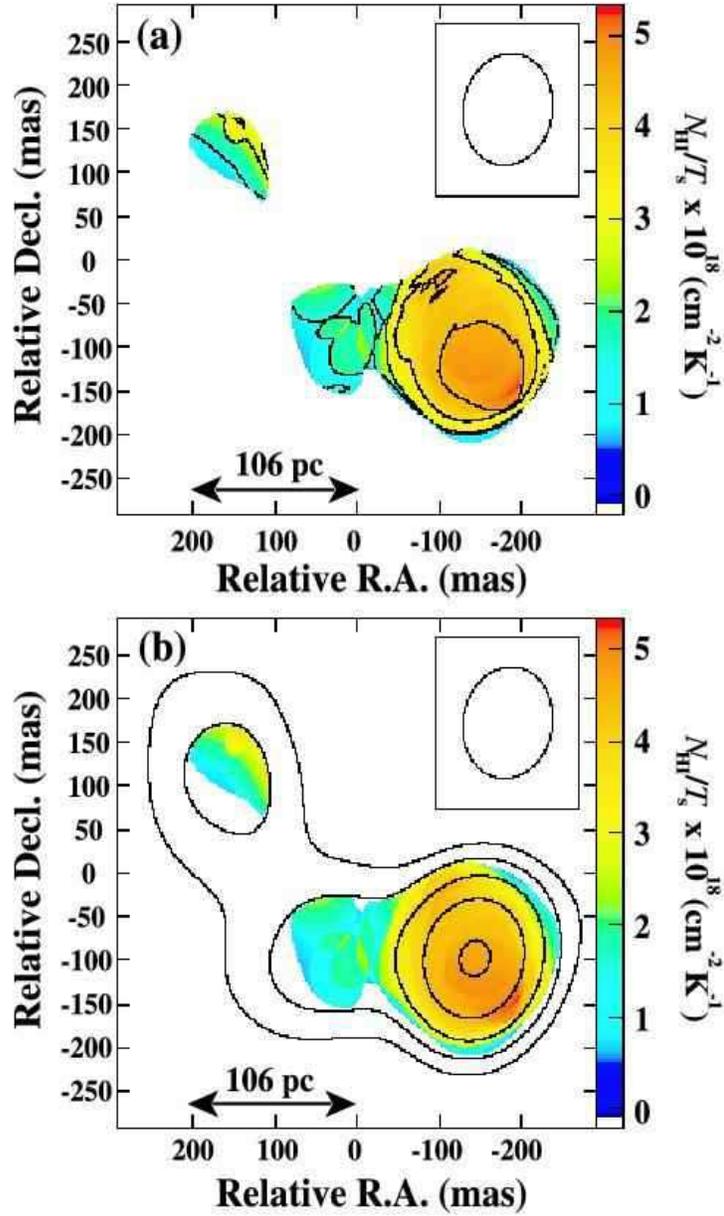}
\caption {$N_{\rm HI}/T_{\rm s}$ images in the velocity range
8626.7--8522.0~km~s$^{-1}$: (a) $N_{\rm HI}/T_{\rm s}$ color image and
contours superimposed; the contours are at 0.5, 1.5, 2.5, 3.5, 4.5 $\times$
10$^{18}$~cm$^{-2}$~K$^{-1}$ (b) $N_{\rm HI}/T_{\rm s}$ color image with
continuum contours superimposed; the contour levels are 3, 6, 12, $\ldots$
48~mJy~beam$^{-1}$. The $129 \times 108$ mas (P.A.~$-13^{\circ}$) restoring
beam is shown in the upper-right corner of each image. The point (0,0) is
$\alpha \rm{(J2000)}=23^{\rm h} 27^{\rm m} 56\rlap{.}^{\rm s} 7195$, $\delta
\rm{(J2000)}=08^{\circ} 46'44\rlap{.}''234$. \label{FIG10}}
\end{figure}

\clearpage
\begin{figure}
\epsscale{0.7}
\plotone{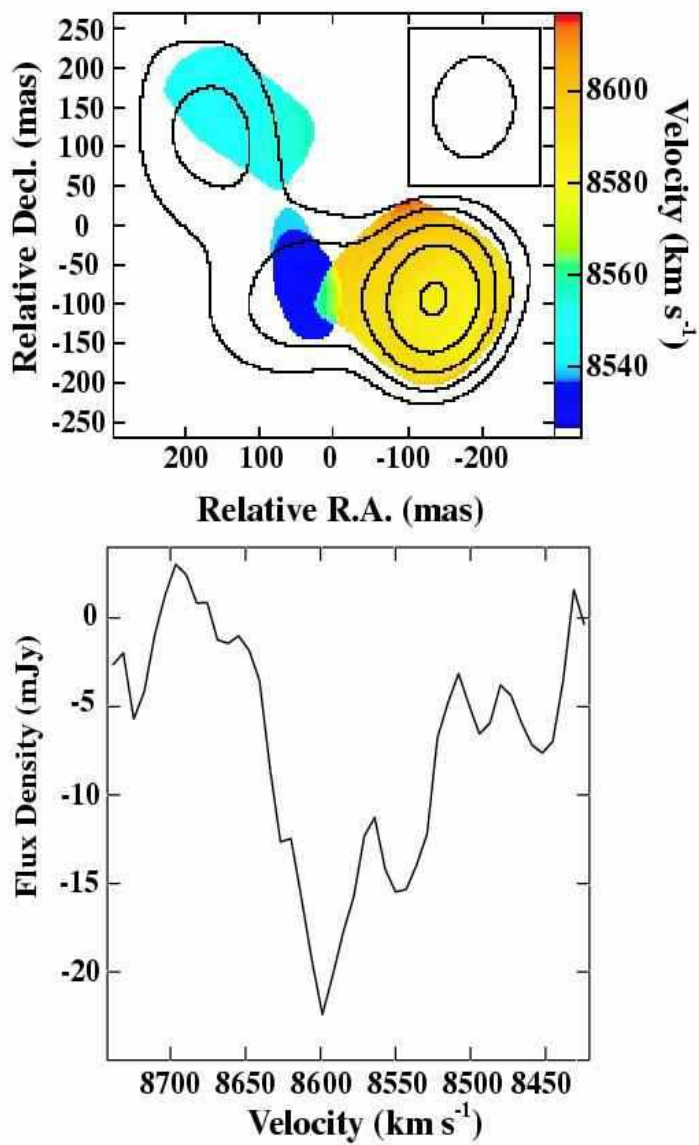}
\caption {{\it Top}: Velocity field of the H~{\footnotesize I}
absorption at low spatial resolution superimposed on the continuum contours
with levels at 3, 6, 12, 24, 48~mJy~beam$^{-1}$. The $129 \times 108$ mas
(P.A.~$-13^{\circ}$) restoring beam is shown in the upper-right corner.
{\it Bottom}: Integrated H~{\footnotesize I} 21~cm
absorption profile against the whole region where H~{\footnotesize I}
absorption is detected at low spatial resolution ($129 \times 108$ mas).
The double H~{\footnotesize I} absorption profile is consistent with the
two absorption features in the H~{\footnotesize I} spectrum obtained with the
305-m Arecibo radio telescope at $3\rlap{.}'3$ resolution
\citep{MIR82}.
\label{FIG11}}
\end{figure}

\clearpage
\begin{deluxetable}{lccc}
\tablecolumns{4}
\tablewidth{0pc}
\tablecaption{P{\footnotesize ARAMETERS} {\footnotesize OF THE} VLBI
O{\footnotesize BSERVATIONS} {\footnotesize OF} NGC~7674}
\tablehead{
\colhead{Parameters} & \colhead{}  & \colhead{} &
\colhead{Values}} \startdata
Observing Date \dotfill & &  & 2001 September~15 \\
Observing Array \dotfill & &  & VLBA + Y27 + AR \\

R.A. (J2000)\dotfill & & &  23 27 56.7097 \\
Dec. (J2000)\dotfill & & &  +08 46 44.135 \\

Total observing time (hr)\dotfill &  & & 12 \\

Phase-referencing cycle time (min)\dotfill &  & &  2.5 \\

Frequency (MHz)\dotfill & & &  1380 \\

Total bandwidth (MHz)\dotfill  & & & 16 \\

High resolution ($11\times 5$~mas) cont.~image rms ($\mu$Jy
beam$^{-1}$)\dotfill  & & & 67\\

Moderate resolution (20~mas) cont.~image rms ($\mu$Jy
beam$^{-1}$)\dotfill  & & & 34\\

Low resolution ($92\times 76$~mas) cont.~image rms ($\mu$Jy
beam$^{-1}$)\dotfill  & & & 78\\

Line velocity resolution (km s$^{-1}$)\dotfill & & & 14 \\

High resolution ($17 \times 5$~mas) line image rms ($\mu$Jy beam$^{-1}$)\dotfill
& & & 260 \\

Low resolution ($129 \times 108$~mas) line image rms ($\mu$Jy
beam$^{-1}$)\dotfill & & & 570 \\

\enddata
\tablecomments{Units of right ascension are hours, minutes, and seconds, and
units of declination are degrees, arcminutes, and arcseconds.}
\end{deluxetable}

\clearpage
\begin{deluxetable}{ccccccccc}
\rotate
\tablecolumns{9}
\tablewidth{0pc}
\tablecaption{G{\footnotesize AUSSIAN} F{\footnotesize ITS} {\footnotesize
TO THE} C{\footnotesize ENTRAL} C{\footnotesize OMPONENT} {\footnotesize IN}
NGC~7674}
\tablehead{
\colhead{} &  & \colhead{Relative Position\tablenotemark{a}}&
\colhead{Peak} & \colhead{Total} &
\colhead{Major Axis\tablenotemark{b}} &\colhead{Minor
Axis\tablenotemark{b}} &\colhead{P.A.\tablenotemark{c}} & \colhead{T$_b$} \\
\colhead{Source} & \colhead{} & \colhead{(mas)} & \colhead{(mJy~beam$^{-1}$)} &
\colhead{(mJy)} & \colhead{(pc)}&\colhead{(pc)}& \colhead{($^{\circ}$)} &
\colhead{(K)} \\ \colhead{(1)} & & \colhead{(2)}& \colhead{(3)} & \colhead{(4)}
& \colhead{(5)} & \colhead{(6)} & \colhead{(7)}& \colhead{(8)}}
\startdata
C1\dotfill & &   0 ,  0  & $16.229 \pm 0.006$&$27.060\pm0.010 $ &$6.4$ &
4.0& 0.16& $21.5 \times 10^7$\\
C2\dotfill & &  7.5W,  0N & $5.023 \pm 0.002$ &$11.783\pm0.006 $ &$7.8$ &
4.7& 5.98& $6.6\times 10^7$\\
C3\dotfill & & 16.5W,  1N & $5.829  \pm0.002 $&$8.405\pm 0.003$  &$6.3$ &
3.5& 178.34& $7.3\times 10^7$\\
\enddata
\tablenotetext{a}{The position (0,0) is $\alpha\rm{(J2000)}=23^{\rm
h} 27^{\rm m} 56\rlap{.}^{\rm s} 7101$, $\delta\rm{(J2000)}=+08^{\circ} 46'
44\rlap{.}'' 134$.}
\tablenotetext{b}{Errors are $\leq 0.02$~pc}
\tablenotetext{c}{Errors are $\leq 0.03^{\circ}$}
\end{deluxetable}

\clearpage
\begin{deluxetable}{ccccccccc}
\rotate
\tablecolumns{9}
\tablewidth{0pc}
\tablecaption{G{\footnotesize AUSSIAN} F{\footnotesize ITS} {\footnotesize
TO THE} W{\footnotesize ESTERN} C{\footnotesize OMPONENT} {\footnotesize IN}
NGC~7674}
\tablehead{
\colhead{} &  & \colhead{Relative Position\tablenotemark{a}}&
\colhead{Peak} & \colhead{Total} &
\colhead{Major Axis\tablenotemark{b}} &\colhead{Minor Axis\tablenotemark{b}}
& \colhead{P.A.\tablenotemark{c}} & \colhead{T$_b$}\\
\colhead{Source} &
\colhead{} & \colhead{(mas)} & \colhead{(mJy~beam$^{-1}$)} & \colhead{(mJy)} &
\colhead{(pc)}&\colhead{(pc)} & \colhead{($^{\circ}$)} & \colhead{(K)}\\
\colhead{(1)} & &
\colhead{(2)}& \colhead{(3)} & \colhead{(4)} & \colhead{(5)} & \colhead{(6)}&
\colhead{(7)}& \colhead{(8)}}
\startdata
W1\dotfill & &     0 ,  0  & $3.229\pm 0.002 $ &$12.386\pm 0.009$ &9.7 & 6.1&
174&$4.3\times 10^7$ \\
W2\dotfill & &   13.5E, 1N & $0.891\pm 0.001$ &$2.231\pm 0.002 $ &7.9 & 4.9&
169&$1.2\times 10^7$ \\
W3\dotfill & &   22.5E, 3N & $1.121\pm 0.002 $&$2.726\pm 0.004 $ &8.9 & 4.2&
175&$1.5\times 10^7$ \\
W4\dotfill & &   34.5E, 6N & $0.628\pm 0.001 $&$2.132\pm  0.002$ &10.0& 5.3 &
164&$0.8\times 10^7$ \\
\enddata \tablenotetext{a}{The position (0,0) is
$\alpha\rm{(J2000)}=23^{\rm h} 27^{\rm m} 56\rlap{.}^{\rm s} 6790$,
$\delta\rm{(J2000)}=+08^{\circ} 46' 44\rlap{.}'' 369$.}
\tablenotetext{b}{Errors are $\leq 0.03$~pc}
\tablenotetext{c}{Errors are $\leq 0.002^{\circ}$}
\end{deluxetable}

\clearpage
\begin{deluxetable}{cccccccc}
\rotate
\tablecolumns{8}
\tablewidth{0pc}
\tablecaption{P{\footnotesize ARAMETERS} {\footnotesize OF THE}
C{\footnotesize ONTINUUM} C{\footnotesize OMPONENTS}
{\footnotesize IN} F{\footnotesize IGURE}~4}
 \tablehead{
\colhead{} &  & \colhead{Relative Position\tablenotemark{a}}&
\colhead{Peak} & \colhead{Total} & \colhead{Major Axis\tablenotemark{b}} &
\colhead{Minor Axis\tablenotemark{b}} & \colhead{P.A.\tablenotemark{c}} \\
\colhead{Source} & \colhead{} &  \colhead{(mas)} &
\colhead{(mJy~beam$^{-1}$)} & \colhead{(mJy)} & \colhead{(pc)}& \colhead{(pc)}&
\colhead{($^{\circ}$)} \\
\colhead{(1)} & &
\colhead{(2)}& \colhead{(3)} & \colhead{(4)} & \colhead{(5)} & \colhead{(6)} &
\colhead{(7)}}
\startdata
C\dotfill & &     0 ,  0  & $49.74 \pm 0.01$ &$52.74 \pm 0.01$&$49.56 \pm
0.02$&$ 41.85 \pm 0.02$& 17\\
E\dotfill & &    168E, 1N & $6.46 \pm  0.02$ &$17.64 \pm 0.05$&$77.29 \pm
0.24$&$ 69.13 \pm 0.32$& 78\\
W\dotfill & &   450W, 226N & $19.71\pm0.04$&$25.07\pm0.04$ &$51.20 \pm 0.10 $&$
48.60 \pm 0.15$&176\\
NE\dotfill & &   317E, 213N & $5.59\pm 0.01$ &$20.57 \pm 0.03$&$99.98 \pm 0.18
$&$ 72.01 \pm 0.17$&30\\
SW\dotfill & &   633W, 68S & $2.45 \pm 0.01 $ &$15.73\pm 0.06$&$144.11 \pm 0.98
$&$87.31 \pm  0.65$&164\\
J\dotfill & &   -  &  - & 7.9 &218& 78 & 116\\
\enddata
\tablenotetext{a}{The
position (0,0) is $\alpha\rm{(J2000)}=23^{\rm h} 27^{\rm m} 56\rlap{.}^{\rm s}
7098$, $\delta\rm{(J2000)}=+08^{\circ} 46' 44\rlap{.}'' 137$.}
\tablenotetext{b}{Half power width for components with gaussian fits. Linear
extent on the plane of the sky for the jet component}
\tablenotetext{c}{Gaussian fit errors are $\leq 0.002^{\circ}$}
\end{deluxetable}

\clearpage
\begin{deluxetable}{ccc}
\tablecolumns{3}
\tablewidth{0pc}
\tablecaption{P{\footnotesize ROPERTIES OF THE} C{\footnotesize ONTINUUM}
 S{\footnotesize
TRUCTURES}}
\tablehead{
\colhead{} & \colhead{Magnetic Field} & \colhead{Pressure} \\
\colhead{Source} & \colhead{(Gauss)} & \colhead{(dyne~cm$^{-2}$)} \\
\colhead{(1)} & \colhead{(2)}& \colhead{(3)}}
\startdata
C\dotfill  & $6.3 \times 10^{-4}$ & $12.3 \times 10^{-9}$ \\
E\dotfill  & $3.0 \times 10^{-4}$ & $2.8 \times 10^{-9}$ \\
W\dotfill  & $3.8 \times 10^{-4}$ & $4.5 \times 10^{-9}$ \\
NE\dotfill & $2.8 \times 10^{-4}$ & $2.5 \times 10^{-9}$ \\
SW\dotfill & $1.6 \times 10^{-4}$ & $0.8 \times 10^{-9}$ \\
\enddata
\end{deluxetable}

\clearpage
\begin{deluxetable}{ccccccccc}
\rotate
\tablecolumns{11}
\tablewidth{0pc}
\tablecaption{G{\footnotesize AUSSIAN} F{\footnotesize ITS} P{\footnotesize
ARAMETERS OF THE} H {\footnotesize I} A{\footnotesize BSORPTION} F{\footnotesize EATURES AT}
H{\footnotesize IGH} R{\footnotesize ESOLUTION}}
\tablehead{
\colhead{} &
\colhead{Velocity} &
\colhead{}  &
\colhead{$\Delta v_{\rm FWHM}$} &
\colhead{} &
\colhead{} &
\colhead{} &
\colhead{$N_{\rm HI}/T_{\rm s}$}
&
\colhead{$N_{\rm HI}$}
\\
\colhead{Feature} &
\colhead{(km s$^{-1}$)} &
\colhead{} &
\colhead{(km s$^{-1}$)}  &
\colhead{} &
\colhead{$\tau_{\rm peak}$} &
\colhead{} &
\colhead{(cm$^{-2}~{\rm K}^{-1}$)}
&
\colhead{(cm$^{-2}$)}
\\
\colhead{(1)} &
\colhead{(2)} &
\colhead{} &
\colhead{(3)} &
\colhead{} &
\colhead{(4)} &
\colhead{} &
\colhead{(5)}
&
\colhead{(6)}
}

\startdata
1 \dotfill & 8603.1~$\pm$~0.7 & & 35.4~$\pm$~2.3& &0.412~$\pm$~0.025&
&(2.8~$\pm$~0.3)$\times{\rm 10}^{19}$
&(2.8~$\pm$~0.3)$\times{\rm 10}^{21}$
\\
2 \dotfill & 8577.6~$\pm$~4.8 & & 98.3~$\pm$~7.1& &0.164~$\pm$~0.028&
&(3.1~$\pm$~0.6)$\times{\rm 10}^{19}$
&(3.1~$\pm$~0.6)$\times{\rm 10}^{21}$
\\
3 \dotfill & 8573.3~$\pm$~1.8 & & 22.5~$\pm$~4.3& &0.298~$\pm$~0.051&
&(1.3~$\pm$~0.4)$\times{\rm 10}^{19}$
&(1.3~$\pm$~0.4)$\times{\rm 10}^{21}$
\\
4 \dotfill & 8534.3~$\pm$~1.2 & & 18.0~$\pm$~2.9& &0.337~$\pm$~0.047&
&(1.2~$\pm$~0.3)$\times{\rm 10}^{19}$
&(1.2~$\pm$~0.3)$\times{\rm 10}^{21}$
\\
\enddata
\end{deluxetable}

\clearpage
\begin{deluxetable}{ccccccccc}
\rotate
\tablecolumns{11}
\tablewidth{0pc}
\tablecaption{G{\footnotesize AUSSIAN} F{\footnotesize IT} P{\footnotesize
ARAMETERS OF THE} H {\footnotesize I} A{\footnotesize BSORPTION} F{\footnotesize EATURES AT}
L{\footnotesize OW} R{\footnotesize ESOLUTION}}
\tablehead{
\colhead{} &
\colhead{Velocity} &
\colhead{}  &
\colhead{$\Delta v_{\rm FWHM}$} &
\colhead{} &
\colhead{} &
\colhead{} &
\colhead{$N_{\rm HI}/T_{\rm s}$}
&
\colhead{$N_{\rm HI}$}
\\
\colhead{Feature} &
\colhead{(km s$^{-1}$)} &
\colhead{} &
\colhead{(km s$^{-1}$)}  &
\colhead{} &
\colhead{$\tau_{\rm peak}$} &
\colhead{} &
\colhead{(cm$^{-2}~{\rm K}^{-1}$)}
&
\colhead{(cm$^{-2}$)}
\\
\colhead{(1)} &
\colhead{(2)} &
\colhead{} &
\colhead{(3)} &
\colhead{} &
\colhead{(4)} &
\colhead{} &
\colhead{(5)}
&
\colhead{(6)}
}

\startdata
I \dotfill & 8600.3~$\pm$~1.1 & & 49.8~$\pm$~2.5& &0.369~$\pm$~0.016&
&(3.5~$\pm$~0.2)$\times{\rm 10}^{19}$
&(3.5~$\pm$~0.3)$\times{\rm 10}^{21}$
\\
II \dotfill & 8550.7~$\pm$~1.8 & & 30.8~$\pm$~4.4& &0.647~$\pm$~0.080&
&(3.8~$\pm$~0.7)$\times{\rm 10}^{19}$
&(3.8~$\pm$~0.6)$\times{\rm 10}^{21}$
\\
III \dotfill & 8544.0~$\pm$~8.9 & & 165.0~$\pm$~14.6& &0.104~$\pm$~0.010&
&(3.3~$\pm$~0.3)$\times{\rm 10}^{19}$
&(3.3~$\pm$~0.3)$\times{\rm 10}^{21}$
\\
IV \dotfill & 8536.3~$\pm$~1.7 & & 22.6~$\pm$~4.0& &0.434~$\pm$~0.067&
&(1.9~$\pm$~0.4)$\times{\rm 10}^{19}$
&(1.9~$\pm$~0.4)$\times{\rm 10}^{21}$
\\
\enddata
\end{deluxetable}

\end{document}